\documentclass{elsart}

\usepackage{epsfig}

\usepackage{amssymb}
\begin{document}

\def\bkg{background}
\def\he4{ ^4\!{\rm He} }
\def\cfo{ $^{14}{\rm C}$ }
\def\c132{ $^{13}{\rm C}/^{12}{\rm C}$ }
\def\carb{ ^{12}\!{\rm C} }
\def\enu{E_{\nu}}
\def\Te{T_e}
\def\angsun{\theta_{\odot}}
\def\be7{^7\!{\rm Be}}
\def\WAYNE{$^a$}
\def\PSU{$^b$}
\def\PITT{$^c$}

\begin{frontmatter}


\title{The chemical history of  $^{14}{\rm C}$ in deep oilfields}
\corauth[cor1]{Physics Dept., Wayne State University, Detroit MI 48201, USA. Tel.: +1-313-577-1444; Fax: +1-313-577-3932. giovanni@physics.wayne.edu}
\author{G. Bonvicini$^a$\corauthref{cor1}} 
\author{N. Harris$^b$, V. Paolone$^c$}
\address{$^a$Wayne State University, Detroit MI 48201}
\address{$^b$Pennsylvania State Univ., University Park, PA 16802}
\address{$^c$Univ. of Pittsburgh, Pittsburgh PA 15260}
%
\begin{abstract}
  14C is an overwhelming background in low-background underground experiments,
to the point where the observation of the all-important (pp) neutrinos from the
Sun can not be observed in carbon-containing experiments. This paper shows that
14C purity can be improved by four orders of magnitude by a careful selection
of the gas field. Two large reduction factors are at work: the low chemical
affinity of methane to single carbon, and the migration of natural gas away
from nitrogen-bearing kerogen during as the oilfield matures.
\end{abstract}
\end{frontmatter}

%
%
%
\maketitle

\section{Introduction.}
All present solar neutrino and dark matter
experiments, and most of the future or planned ones, 
use carbon-containing materials within the fiducial volume. In some cases,
such as the current BOREXINO and KAMLAND detectors, the \cfo beta decay
(with an end point of 156 keV) background in hundreds of tons of organic 
scintillator
is so overwhelming
as to prevent observation of the all-important
$(pp)$ solar neutrino flux (with an end point of 217 keV).

In the case of the experiment we propose,
the Solar Neutrino TPC,\footnote{TPC means Time Projection Chamber.}
methane is about 3\% by volume and 12\% by mass\cite{TPC}. 
Because the TPC is directional, and because only a fraction of the 
fiducial mass is carbon, the TPC tolerance to \cfo impurities is at least
three orders of magnitude above BOREXINO\cite{borexino} and KAMLAND. Nonetheless a
preliminary study indicates that \cfo might
be a problem for the TPC\cite{TPC}, and that a \cfo reduction by
a factor of ten or more would be desirable.

In view of these problems it is worth studying ways to reduce the content of
\cfo of our experiments, which is defined as
\[ r=^{14}{\rm C}/^{12}{\rm C}.\]

In the process we solve a puzzle related
to low background physics. 
The BOREXINO Collaboration analyzed the \cfo content of methane
before embarking in the construction of the large detector\cite{toronto}. 
They found the limit $r<10^{-18}$. However, once they built their first 
multi-ton prototype they found that, in their scintillator, 
$r=(1.94\pm0.09)10^{-18}$\cite{borexino}.
A recent measurement of another batch of scintillator, produced with 
petroleum from a different oilfield, gave another result,  
$r=11\times 10^{-18}$\cite{schoenert}. From the work described below, we 
conclude that the apparent discrepancy is most likely due to the fact
that the former was methane, and the latter scintillator, and they have
vastly different \cfo content.

Schoenert and Resconi\cite{schoenert} have provided a first understanding
of the $r$ value in petroleum, and this work builds on their. They have 
identified the $^{14}$N(n,p)\cfo reaction as the main source of \cfo, 
they have modeled the dependence of $r$ on the uranium content of the rock
and the nitrogen content of the petroleum, and they have predicted 
a $r$ range of $10^{-16}-10^{-20}$. Section 2 is mostly a description of
their work. 

In this paper the steps through which
\cfo is originated and incorporated in molecules are retraced. 
All the carbon
in use in our experiments was originally extracted from oilfields, and
petroleum provided the industrial 
feedstock for the production of various plastics and
scintillating liquids in use in low energy experiments.
Our main result is that, if petroleum is carefully chosen, $r$ may go down
to the $10^{-22}$ level, making $(pp)$ solar neutrino experimentation at 
BOREXINO
and KAMLAND a possibility. It is also all but certain 
that the Solar Neutrino TPC will have negligible  \cfo backgrounds.

Section 2 summarizes the relevant information about geological strata, 
oilfield parameters, and underground nuclear interactions. Section 2
discusses also the first high precision \cfo measurement, by 
BOREXINO\cite{borexino}, and the likeliest possible explanation of the
results. 
Section 3
briefly discusses isotopic separation, and provides experimental tables
about the chemical reactivity of single carbon and hydrocarbons. 
Section 4 lists
possible pitfalls, suggests a
set of inexpensive measurements, and offers some criteria to select low
\cfo hydrocarbons.

\section{What we know.}
\subsection{Oilfields, gasfields and surrounding rock.}
Oil and gas fields are found throughout the world in a wide
variety of geological
formations, over a range of depths from near the surface to 10000 meters.
Much more rarely, hydrocarbons are encountered in metamorphic or igneous 
rocks. The most common types of sedimentary rocks hosting oil and gas 
pools are
sandstone (predominantly SiO$_2$) and
limestone (predominantly CaCO$_3$). Other lithologies
such as shale (clay) are less common. 
Oil and natural gas are 
found in rock pores, whose typical size ranges from 0.1$\mu$m
to 1mm. The petroleum 
content of the rock is typically 0.05-0.3 by volume. 

Because petroleum originates from decayed biomass,
it is supposed to contain far more 
nitrogen (via protein decomposition) than typical crustal rock. Nitrogen
content of oil averages 2\%,
with variations 
depending on whether
the organic matter was derived from
plants, algae, or bacteria.
The typical 
nitrogen content of crustal rock, in contrast, is a few tens of 
ppm\cite{ore}.

Oil and gas generally coexist with water, although occurrences in dry rock
are recorded. Water occurs both as thin films adhering to grain surfaces
and as discrete lenses within the hydrocarbon volume.
Hydrogen sulfide
(H$_2$S) is a common minor constituent of natural gas, as are carbon 
dioxide(CO$_2$) and nitrogen (N$_2$). In rare instances, these
compounds can dominate the composition of natural gas.

Reservoirs occur at temperatures of 280 to 460K, a function of the reservoir
depth and the local geothermal gradient. Reservoir pressures range up to 
$10^4$ atmospheres, corresponding to the weight of the overlying water column
(hydrostatic pressure), the overlying rock column (lithostatic pressure) or
intermediate values.
Because Archimedes forces act on the petroleum, oil and gas
tend to separate in depth 
within the same oilfield, although some gas typically
dissolves in coexisting oil. Also, Archimedes forces may move
petroleum laterally as it tries to move upward, with migrations of tens 
of kilometers possible.

The composition of most natural
gas is dominated by  methane, comprising more than 80\% of the gas by
volume. Some gases have been described with compositions highly depleted
in methane and enriched in non-hydrocarbon gases such as nitrogen,
helium, H$_2$S and CO$_2$. Oil
has a very broad hydrocarbon spectrum. 

In many cases considerable amounts of data exist about oil
and gas fields, including
well preserved samples of petroleum and host rock, and  their
chemical composition, temperature, and pressure. Most oil and gas 
fields are also
analyzed by lowering sensing tools on electric cable down boreholes
(logging tools), in order to obtain information on the presence, type and
concentration of hydrocarbons, the mineral composition of the rock, the amount
of pore space and the presence of fractures.

\subsection{Nuclear reactions in the rock and petroleum.}
In the upper atmosphere, and all the way down to 50mwe below ground, the 
neutron flux (which, we will find later, ultimately 
controls the production of \cfo)
is essentially generated by downward cosmic rays\cite{nolte}.
At depths exceeding 300mwe however the cosmic flux has abated to
negligible levels, and the reactions that ultimately produce \cfo
are dominated by nuclear reactions in the rock.

There are five reactions of interest which need to be 
considered\cite{borexino},
listed in Table 1. Their ordering represents the expected order of importance
in typical crustal rock\cite{borexino}.
\begin{table}
\caption{Reactions that produce \cfo in an oilfield.
\label{tn:resol}}
\begin{center}
\begin{tabular}[t]{|c|c|c|}
\hline
  Reaction   & $\sigma$(millibarn) & Comment\\
 \hline
$^{17}$O(n,$\alpha$)\cfo & 235 & - \\
$^{14}$N(n,p)\cfo & 1830 & - \\
$^{13}$C(n,$\gamma$)\cfo & 1.4 & - \\
$^{11}$B($\alpha$,n)\cfo & 20-200 & 2-5 MeV energy \\
 $^{226}$Ra$\rightarrow ^{14}$C & - & Tripartition \\

\hline
\end{tabular}
\end{center}
\end{table}

$\alpha$ particles in crustal rock
are predominantly generated by the decays of $^{238}$U
and  $^{232}$Th, and their daughters. Neutrons are predominantly generated
by ($\alpha$,n) reactions in other elelents in the rock 
(such as Al, Mg, or Na). Because neutrons are ultimately generated by
$\alpha$ particles,
the neutron flux is ultimately controlled
by Uranium and Thorium concentrations
($^{226}$Ra is a $^{238}$U daughter as well). 

The range of neutrons, of course, far exceeds the size of the rock pores.
The range of 1 MeV $\alpha$ particles is of order tens of $\mu$m.
In all
the reactions discussed above the \cfo is imparted an energy of order 100 
keV or larger, 
which results in a typical range of order one micron in rock, 
few $\mu$m in oil, and
a fraction of a millimeter in gas, to be compared with pore sizes varying from
0.1$\mu$m to 1 mm. 

\subsection{The BOREXINO result and its interpretation.}

BOREXINO\cite{borexino} 
eventually measured $r$ in their scintillator to 
unprecedented accuracy and found it to be $(1.94\pm0.09)10^{-18}$. 
The original paper advances
the hypothesis that the sample was contaminated by atmospheric carbon
during production\cite{borexino}.

\begin{table}
\caption{Average and range U, 
Th, B content in typical oilfield rock.
The neutron flux $n$, last column, is defined in Sect. 2.3 and a
order of magnitude estimate only is given.
\label{tn:res}}
\begin{center}
\begin{tabular}[t]{|c|c|c|c|c|}
\hline
  Type of rock   & U(ppm) & Th(ppm) & B(ppm) &$n$ (n/grams/year)\\
     & Refs.\cite{schoenert,turekian,krauskopf} & Refs.\cite{schoenert,turekian,krauskopf} & Ref.\cite{harder}& Ref.\cite{schoenert}\\
 \hline
Sandstone & 0.45 (0.2-0.6) & 1.7 (0.7-2.0) & 30 (5-70) & 1. \\
Limestone & 2.2 (0.1-9.0) & 1.7 (0.1-7.0) &20 (2-95) & 1. \\
Shale & 3.5 & 11 & 130 (25-800)  & 1. \\

\hline
\end{tabular}
\end{center}
\end{table}

Schoenert and Resconi\cite{schoenert} first 
pointed out that the nitrogen content of oilfields
is much higher than normal crustal rock (a nominal content of 5\% is chosen
in their analysis\cite{schoenert}, a factor of 2.5 higher than that in 
Ref.\cite{hunt}), 
and that as a result the reaction
 $^{14}$N(n,p)\cfo is the dominant one in oilfields. 
If that is the case, then $r$ in petroleum is
regulated by three main factors:
\begin{itemize} 
\item the neutron emission intensity
\item the neutron absorption by nuclei other than $^{14}$N
\item the $^{14}$N concentration
\end{itemize}

The intensity is clearly regulated by the $^{238}$U concentration in the
rock. $^{232}$Th is also a major contributor, however the concentration
of these two isotopes tends to be correlated most everywhere in geological 
formations, so that only the $^{238}$U concentration needs to be tracked.
This concentration is defined as $U$. The dependence on $U$ is model
independent, if $U=0$ then $r$ will be extremely small. 
The $^{238}$U
content varies in a way familiar to low background physicists. 
Shale is known to have the highest
uranium  content of the three kind of strata considered here. Therefore
obtaining petroleum from a limestone formation would contribute one order
of magnitude.

The intensity is also regulated by the concentration of the various nuclei
which ultimately generate the neutrons via ($\alpha$,n), weighted by the
relative cross section. This concentration
is defined as $M$. This parameter regulates $r$ only under the hypothesis
that the predominant reaction is initiated by a neutron. $M$ is made up of a 
long list of nuclides, which makes it a hard parameter to control. In view of
our main problem, finding specifications for the oilfield with low  \cfo,
we have not sought a characterization of this parameter.

The neutron absorption will be dominated under most circumstances by the
presence of boron concentrations (typically in the few tens ppm - few hundred
ppm in crustal rock). This concentration
is defined as $B$. Note that boron 
will lower $r$ only as long
as the predominat reaction is initiated by a neutron. In the case where
the predominant reaction is $^{11}$B($\alpha$,n)\cfo, Boron will
increase $r$. Boron content varies by more than two orders of magnitude
in crustal rock (see Table 2).

Finally, the N concentration is defined as $N$, 
and regulates $r$ only
if the reaction $^{14}$N(n,p)\cfo is dominant. The nitrogen content
of the rock may vary by a factor of a few, primarily depending on the
shale content of the reservoir, since nitrogen content in rock is primarily
associated with clay minerals.

We also identify
$UM/B$ as $n$, the total neutron flux.
$r$ is then
roughly proportional to
\[ r \propto {UMN\over B}=nN.\]

The formula above provides a starting point to try and reduce 
$r$ by
selecting the appropriate oilfield as the feedstock source. Minimization
of the ratio above will minimize the $^{14}C/^{12}C$ ratio for 
petroleum in bulk.

\section{The chemical history.}
The previous Section described how to select the best possible OILFIELD.
In this Section we describe how to select the best possible HYDROCARBON.
It is also important to notice that this Section is totally unrelated to 
the previous one. Independent of the nuclear reaction that produces \cfo , 
the nucleus starts its chemical history as a single carbon atom.

We concentrate on methane (CH$_4$), the gas we chose as the quencher in the 
TPC\cite{TPC}.
There are several reasons to look at methane. It can be trivially separated
from all other hydrocarbons by boiling off at 112K (ethane boiling point is
at 184K), 
it can be easily stripped of 
extra hydrogen atoms by industrial methods, to manufacture a different
hydrocarbon to be used as scintillator 
feedstock, and isotopically pure methane is
commmercially available. Its chemistry is different from that of other 
hydrocarbons, and it is the only hydrocarbon that can be extracted 
almost unmixed from a gas field.

\subsection{Isotopical separation.} 
Commercial quantities of isotopically pure (\c132 ratio of $10^{-5}$ to 
$10^{-6}$, as opposed to the natural \c132 ratio of $10^{-2}$) 
methane are available, and
are used to produce chemical diamond heat sinks with extreme heat 
conductivity. The typical cost is of order 
\$10-\$50 per gram for gram quantities (depending on purity),
and while the cost would come down dramatically for mass production it is
clear that the cost of kilotons of isotopically pure scintillator is 
prohibitive. It remains a viable possibility where kilograms of isotopically
pure feedstock are needed.

The best current technology available for carbon isotopic enrichment makes use
of carbon monoxide (CO)\cite{ishida}, 
where the heavier carbon nucleus changes the
interatomic distance appreciably and makes separation easiest. If the ratio 
\c132
is decreased by a factor K during enrichment, then $r$ is
decreased by a factor 2K. A plant that could produce large amounts of 
isotopically pure hydrocarbon is fairly complicated, and includes the
transformation of CO into the desired feedstock, plus the disposal of large
amounts of waste products.

\subsection{Single carbon chemistry.}
As stated above, no matter the nuclear reaction that produces the \cfo,
one always starts with a single carbon atom, with an energy of order 100 keV,
eventually thermalizing in the medium and interacting to form a molecule.

A wealth of single carbon
chemical data, interacting with dozens of hydrocarbons,
were obtained in the 1960s and are reproduced here. Ref.\cite{revie},
for example, cites over twenty works where the chemical history of radioactive
single carbon in hydrocarbon mixtures was traced.
Much of the data were
obtained using $^{11}$C isotopes, produced by irradiating the hydrocarbons
with neutrons
($^{12}$C(n,2n) $^{11}$C).
This has the advantage of including any ``hot chemistry'' effect
in the measurements, as the $^{11}$C isotope had to slow down and stop in the
hydrocarbon medium, and in the process possibly 
interact with free radicals and/or excited molecules present in its wake. 
Indeed some of the data
below suggest that ``hot chemistry'' is at work at the percent level. The
chemistry of  $^{11}$C and \cfo are assumed to be the same for 
the precision
of this work. 

It is clear that methane is difficult to produce,
starting from single carbon atoms in a hydrocarbon medium. 
Consider the following chain of reactions:
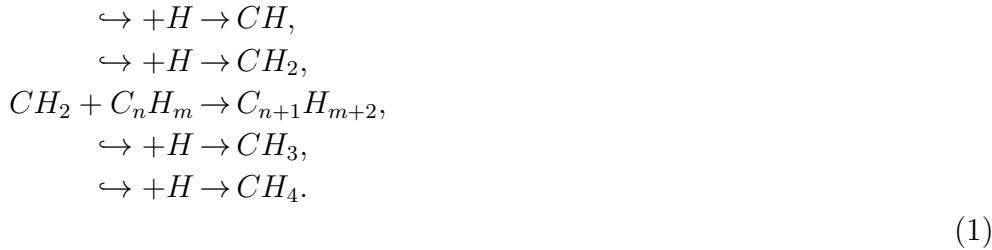
\begin{eqnarray}
C +C_nH_m &\rightarrow & C_{n+1}H_m,\nonumber \\
\hookrightarrow +H&\rightarrow & CH,\nonumber \\
\hookrightarrow +H&\rightarrow & CH_2,\nonumber \\
CH_2 +C_nH_m &\rightarrow & C_{n+1}H_{m+2},\nonumber \\
\hookrightarrow +H&\rightarrow & CH_3,\nonumber \\
\hookrightarrow +H&\rightarrow & CH_4.\nonumber \\
\end{eqnarray}
The formation of methane must proceed through four ``hydrogen stripping''
steps. If the hydrocarbon is pure methane, there are two branch-off points,
the capture of single carbon to produce ethylene $(CH_4+C\rightarrow C_2H_4)$,
and the capture  of the $CH_2$ radical by methane to produce 
ethane $(CH_4+CH_2\rightarrow C_2H_6)$. In both cases the single carbon 
is sequestered in hydrocarbons that can be trivially separated by boil-off
from methane.

The chemical cross section for ``hydrogen stripping'' tends to be small, 
compared with other processes.
If H$_2$O or H$_2$S are present in quantity, though,
they may act as donors of less strongly bound hydrogen atoms.

The single carbon, and the three intermediate radicals CH, CH$_2$, and CH$_3$,
are very reactive with lifetimes in a liquid 
hydrocarbon medium of order milliseconds. Over that time, they will collide
hundreds of millions to billions of times with the surrounding molecules. 
Traces of compounds with high reactivity will drastically 
alter the outcome of an irradiation test.

In Table 3 the distribution of radioactive hydrocarbons is shown when
methane, and methane mixed with other chemicals, are irradiated to produce
$^{11}$C. There are several points of interest here. First, pure methane 
reduces $r$ by roughly a factor of 7. Second, the last four columns show
that small amounts of other chemicals drastically alter the results. The
reactivity of \cfo with C$_2$H$_4$ is about a million times higher than 
with methane.
Third, the interaction with oxygen donors may produce sizeable amounts of 
$^{14}$CO, which boils at a temperature lower that that of methane 
(82K as opposed to 111K). A double boiloff procedure is called for, where 
first natural gas is boiled off at 112K, leaving behind the higher 
hydrocarbons, and then CO is boiled off at approximately 105K, leaving 
the methane behind. We note that Table 5 below 
lists some gasfields with sizeable
molecular oxygen content. It is not clear at this time if water can act as
an oxygen donor.

\begin{table}
\caption{Distribution of radioactive molecules when methane is irradiated.
All values in percent. Ref.$[14]$ had a typical sensitivity of 0.2\% and
errors of order 1\%. 
Ref.$[15]$ had a typical sensitivity of 2\%.
\label{tn:re}}
\begin{center}
\begin{tabular}[t]{|c|c|c|c|c|c|}
\hline
 Contaminants  & None  & 0.12\% O$_2$  & 2\% O$_2$ & 2\% C$_2$H$_2$ & 1.2\% C$_2$H$_4$\\
    &  (Ref. 14) & (Ref. 14) & (Ref. 15) & (Ref. 14) & (Ref. 14)\\
 \hline
CO        &   $<$0.2  & 20.4 & 26.8 & $<$0.2  &$<$0.2  \\
CH$_4$    &   13.9  & 1.5 & $<$0.15 & 1.9  & $<$0.2  \\
C$_2$H$_2$    &   17.7  & 30.0 & 32.3 & 32.8  & 25.2  \\
C$_2$H$_4$    &   12.4  & 28.0 & 30.5 & 29.5  & 23.5  \\
C$_2$H$_6$    &   23.9  & 3.1 & $<$0.5 & 6.0  & 3.4  \\
C$_3$H$_8$    &   11.1  & NA & $<$0.2 & $<$0.2  & 1.7  \\
Higher    &   20.9  & NA & 16 & 29.8  & 46.2  \\
\hline
\end{tabular}
\end{center}
\end{table}

Table 4 shows the temperature and/or phase dependence of the chemical 
history of single carbon. While temperature and phase effects are
clear, they only affect concentrations by factors of a few.
Dose effects were clearly observed\cite{t13}, and they are likely to be 
responsible for small discrepancies that can be found between different
experiments. Ref.\cite{t13} lists a number
of zero-irradiation percentages extrapolated from runs at different
levels of irradiations. Irradiation dependence does not affect
the precision of the conclusions below.
\begin{table}
\caption{Radioactive methane yield for various hydrocarbons, at various
temperatures, phases, and in presence of contaminants. 
\label{tn:reee}}
\begin{center}
\begin{tabular}[t]{|c|c|c|c|c|c|}
\hline
 Hydrocarbon  & CH$_4$ Yield   & Temp. & Phase & Contam.&Ref. \\
   & (\% )  & (K)  &  &  &  \\
 \hline
CH$_4$        &   13.9   & 298 & Gas    & none      & 14  \\
CH$_4$        &   4.5    & 77  & Solid  & none      & 14  \\
C$_2$H$_6$    &   2.1    & 298 & Gas    & none      & 16  \\
C$_2$H$_6$    &   $<$2.0 & 298 & Gas    & 2\% O$_2$ & 15  \\
C$_2$H$_6$    &   3.7    & 77  & Liquid & none      & 16  \\
C$_3$H$_8$    &   2.9    & 298 & Gas    &none       & 16  \\
C$_3$H$_8$    &   0.5    & 298 & Gas    &0.58\% O$_2$  & 16  \\
C$_3$H$_8$    &   $<$2.0 & 298 & Gas    & 2\% O$_2$ & 16  \\
C$_3$H$_8$    &   $<$0.5 & 298 & Gas    &14\% O$_2$ & 16  \\
C$_3$H$_8$    &   3.4    & 95  & Liquid & None      & 16 \\
C$_3$H$_8$    &   4.7    &107  & Solid  & None & 16 \\
Cyclopropane    &   $<$2.0   & 298 & Gas    & 2\% O$_2$ & 14  \\
Cyclohexane    &   6.7   & 303 & Gas    & none & 17  \\
Cyclohexane    &   5.3   & 273 & Gas    & none & 17  \\
Cyclohexane    &   5.7   &95 & Solid    & none & 17  \\
n-hexane    &   5.6   & 303 & Gas    & none & 17  \\

\hline
\end{tabular}
\end{center}
\end{table}

Generally, in Table 4, methane yields are of order 1-10\% of the 
original sample, and they are 
smallest for smaller hydrocarbons, such as ethane, which is identified as the
best \cfo  absorber amongst the alkanes. Heavier hydrocarbons
seem to cluster around 5-7\%. If all  \cfo is processed
through interactions with C$_2$H$_6$, the \cfo content of methane
will be 2\% of the original irradiation.
Based on Table 3, 
we make the assumption that other hydrocarbons have much higher reactivity.
Based on Table 5 below, 
we assume that other hydrocarbons in a gasfield are going
to be primarily
C$_2$H$_6$ and C$_3$H$_8$. Under these assumptions,  and the measured
branchings in Table 4,
the \cfo content of methane is reduced by a factor between 35 and 48 in
a mixture with ethane and propane.

We conclude that in natural gas the isotopic distribution of methane is
radically different from that of the other hydrocarbons present in the mix,
which is one of the two main results of this work.
The situation of oil is a lot 
more complicated. If there was no methane originally, then all the 
methane will contain \cfo, as it is produced by the chemical reactions
of Table 4. This is a problem that does not concern us. In our TPC, we plan
to use boiled-off natural gas, and we expect it from the data above to
be practically free of \cfo.

\section{Putting it all together.}
Natural gas is formed by the maturation (chemical breakdown) of organic matter
in shales and shaley limestones, followed by
migration to and accumulation in reservoirs. 
In most cases, the concentration of nitrogen in the natural gas is considerably
lower than in the original organic matter due to chemical fractionation very
early in the maturation process. Rarely, nitrogen is subsequently concentrated
in natural gas by chemical reactions that destroy methane and other hydrocarbon
gases.
The large variance of nitrogen content (Table 5)
reminds us that there are other sources of nitrogen underground,
including atmospheric and non-biogenic sources, which may mix with natural gas.

\begin{table}
\caption{Gasfield location, depth, and chemical composition of the gas
in southwestern US and northwestern Mexico.
Data from Refs.$[19]-[24]$.
\label{tn:reee}}
\begin{center}
\begin{tabular}[t]{|c|c|c|c|c|c|c|c|}
\hline
 Location  & Depth(ft)& CH$_4$\%  & C$_2$H$_6$\%& C$_3$H$_8$\% & N$_2$\%& O$_2$\% &H$_2$S\%\\
 \hline
Solano Co, CA       & 4700 & 96.9   & 1.9   & -    & 1.2  & -    & -\\
Solano Co, CA       & 4500 & 94.6   & 2.7   & -    & 1.86 & -    &-\\
Contra Costa Co, CA & 3500 & 89.523 & 4.213 & 2.04 & 1.85 & 0.03 & -\\
Kern Co, CA         & 2160  & 99.77& 0.02   & -     & 0.16& -    &  -\\  
San Joaquin Val, CA & 4300 & 98     & 1.1   & -    & 0.6  &-     &-\\
San Joaquin Val, CA & 2500 & 94     & -     & -    & 4.65 &-     &-\\
San Joaquin Val, CA & 8000 & 43.55  & -     & -    & 55.76&-     & -\\  
San Joaquin Val, CA & 8500 & 81.1   & 8     & 4.56 & 0    &-     &-\\
San Joaquin Val, CA & 5500 & 99.33  & 0.38  & 0.16 & -    &-     &-\\
San Joaquin Val, CA & 7500 & 92.3   & 4.2   & 2.27 & -    &-     &-\\
Ventura Co, CA      & 4600 & 99.55  & 0.15  & 0.11 & -    &-     &-\\
Unita Basin, UT     & 5050 & 93.8   & 3.72  & 1.08 & 0.33 &-     &-\\
Unita Basin, UT     & 4600 & 98.58  & 0.37  & 0.07 &  -   &-     &-\\
Unita Basin, UT     & 2875 & 9.3    & 5.91  & 2.01 & 0.42 &-     &-\\
Tampico Em., Mexico & -    & 86.47  & 4.95  & 3.49 & -    & -    & 1.79\\
Tampico Em., Mexico & -    & 70.92  & 8.51  & 5.78 & -    &-     & 5.32\\
Tampico Em., Mexico & -    & 67.19  & 11.48 & 8.67 & -    &-     & 2.77\\

\hline
\end{tabular}
\end{center}
\end{table}

We now rewrite the formula for $r$ as
\begin{equation}
r \propto n(N+\epsilon_{rock})\epsilon_{CH4}.
\end{equation}

$\epsilon_{rock}$ is the natural activity when no nitrogen is present
(due to the other reactions in Table 1).The product  $n\epsilon_{rock}$ 
is assumed to be 
$\sim 5\times 10^{-21}$\cite{borexino}.  
$\epsilon_{CH4}$ depends on the particular gas 
mixture, but under the hypothesis that ethane and propane dominate it is
0.02 to 0.03.
When nitrogen is absent (e.g., the 8th row in Table 5),
\begin{equation}
r \propto n\epsilon_{rock}\epsilon_{CH4}\sim 1-2\times 10^{-22} .
\end{equation}

This a level at which BOREXINO and KAMLAND may be able to work.

\subsection{Possible pitfalls.}
While it is clear that \cfo does not end easily in methane,
there are two possible ways for the isotope to later enter the 
methane component. They
are discussed here.
\begin{itemize}
\item Bacterial activity. Vast amounts of bacteria 
live underground, and they produce methane
as a byproduct of their methabolic activity. If they are present in the
gas field, they may reintroduce \cfo in methane. Ref.\cite{hunt}
states that bacteria can exist in gas fields only if the temperature
does not exceed 350K. 
\item Equilibrium reactions. For the selection method proposed here to work,
\cfo must stay in the hydrocarbon where it ended at the
end of its chemical history for six half-lives or more (35000 years),
much less than the average residence time of gas in a reservoir. 
If equilibrium reactions (e.g., 
a +C$_2$H$_4\rightleftharpoons$ b+ CH$_4$, with $a$ and $b$ any other
chemical) are significant,
then \cfo will have a way to redistribute itself uniformly across the
hydrocarbon spectrum. These reactions do not happen at 
temperatures of a few hundred degrees Kelvin, witness the clear absence of 
alkenes in natural gas.
The reactions do
happen readily in the laboratory 
at room temperature in the presence of molecular hydrogen,
which needs to be avoided at all costs. 
\end{itemize}
\subsection{A simple test.}
Because chemical selection does not depend on the initial nuclear process, 
a simple and inexpensive apparatus could provide definitive proof of the
arguments above. As long as the two pitfalls above are avoided, this test
has no meaning for the TPC\cite{TPC} - the \cfo backgrounds from methane
will be negligible no matter what. The test, however, would have ramifications
for BOREXINO or KAMLAND, who may need three or four orders of magnitude
suppression of \cfo to be able to study $(pp)$ neutrinos.

A pressure vessel would have to be filled by a mixture of materials that
simulate an underground well (e.g., sand, salt water, oil or gas, and some
nitrogen), and irradiated by a neutron source. The hydrocarbons can then be
separated by boiloff and 
analyzed for \cfo content. This test should be able to determine all the
coefficients of Eq.(2), and it would also test the one assumption we
made, that the single carbon reactivity in natural gas
is dominated by non-methane compounds.
We note that oil research companies, such as
Schlumberger Doll, have such vessels readily available for use\cite{loomis},
so that only the liquids and gases, a boiloff apparatus, and a \cfo 
counting device need to be provided.

\section{Outlook.}
In this work we have advanced the work of Schoenert and Resconi in three
important ways. First, we have identified that there is a chemical
selection at work that depletes the \cfo content of methane (that is, we have
introduced $\epsilon_{CH4}$ in Eq. (2)). Therefore,
we have also identified the best petroleum as gas. Second, we note
that nitrogen becomes more easily separated from gas than from oil,
as shown in Table 5 (we minimize $N$ in Eq. (2)). 
Third, because hydrogen stripping is identified
as the primary chemical reaction for \cfo to enter methane, we 
can better specify (or test in the future, Sect. 4.2) 
the properties of a gas field that is depleted in \cfo .

Several things need to be stressed before giving the specifications for
the perfect gasfield. The result of Eq. (3) is to be understood as an
order of magnitude estimate. Variations in the $M$ and $B$ parameters
may be at that level, and may not generally be known.
We also note that the $U$ parameter is generally
not known for many oilfields, whereas the gas composition is 
generally well known (Table 5). There are practical advantages in picking
a gas field based on easily measurable quantities (just the gas composition). 

We stress once again that isotopic purification by chemical means works
regardless of the nuclear source of \cfo. Purification
by nitrogen content will work if nitrogen is the primary source of \cfo. 

\subsection{The perfect gasfield.}
It is clear from the discussion that only a gasfield should be considered, if 
large reductions in  \cfo are sought. Here we list the properties of the
gas field. The first three are meant to minimize the values of Eq.(2), whereas
the last four are meant to avoid \cfo migrating back into the
petroleum, avoid hydrogen donors, atmospheric contamination, and carbon
monoxide. The first two conditions were given first by Schoenert and 
Resconi\cite{schoenert}.

{\bf Low nitrogen content.} Gaseous nitrogen content is readily measured 
and can be very low. Besides the gas fraction that is nitrogen, some attention
is to be given to the potentially 
nitrogen-rich shales that often form the seal
rock above a reservoir. For the TPC purposes, a nitrogen content not exceeding
10\% (most gasfields) is adequate. For BOREXINO or KAMLAND, the nitrogen 
content should be below or of order 1\%. Two orders of magnitude \cfo reduction
are available by simply requiring low nitrogen.

{\bf Proven low uranium content.} This is the cut that will reduce the
number of candidate gasfields. Only the U/Th content of a number of 
them is known. Note that this condition is not necessary for the TPC,
gas analysis alone should provide sufficiently pure methane.

{\bf Methane/ethane/propane dominated mixture.} The vast majority of gasfields
satisfy this requirement, so it is hardly worth mentioning. This data
is easiest to obtain. Almost two orders of magnitude decrease in $r$ are 
available here (Tables 3 and 4).
Under all circumstances the reduction should be at least one order of 
magnitude (Table 3). 

{\bf Minimum water.} It is not clear that the chemical and nuclear
properties of water will ultimately contribute an
increase in $r$. On the positive side, underground water can be 
extremely saline, chlorine is
a strong neutron absorber (similar to Boron, once their concentrations
are taken into account)
and water is an oxygen donor, which helps 
sequester \cfo (Table 4). On the negative side,
water may or may not 
be an hydrogen donor, water of meteoric origin may be present in the gasfield
and carry atmospheric CO$_2$, which has $r=10^{-12}$. A further
layer of complexity is introduced by having water in film or lens form. 
This is one of the major reasons to go through 
the testing procedure described
in Section 4.2. The biological reason for avoiding water is that 
if there is no water, there can not be any bacterial activity. 

{\bf No bacterial activity.} This specification can be enforced by requiring 
the reservoir temperature to be at least 350K.

{\bf No H$_2$S.} The most loosely bound hydrogen underground belongs to H$_2$S,
and so it needs to be avoided. We note that this parameter is also easily
measured, and that H$_2$S can be absorbed by the rock if iron is present.

{\bf Double boiloff.} All other chemical species that may carry \cfo 
boil at temperatures well above methane, except CO. A second boiloff should 
be performed to vent any possible CO traces before use.

{\bf Other considerations.} It is possible that oxygen donors, such as oxygen
and carbon dioxide, will help suppress \cfo (Tables 3 and 5). 
On the other hand,
it is possible that these compounds be of atmospheric origin. 
Unless the reservoir is deep enough to exclude atmospheric contamination,
it is best to require low levels of any non-hydrocarbon gas. The effect of 
these compounds can be tested using the setup of Sect. 4.2.

\subsection{Acknowledgements.}
We wish to than T. Ishida, J. Kadyk, T. Loomis, O. Mullins, 
R. Raghavan, K. Sauer, S. Schoenert, 
P. Shevlin and J. Zumberge
for useful discussions.


\clearpage

\begin{thebibliography}{99}

\bibitem{TPC} G. Bonvicini, D. Naples, V. Paolone, hep-ex/0109032,
submitted to Nuclear Instruments and Methods.
\bibitem{borexino} G. Alimonti {\it et al.}, Phys. Lett. B 422: 349, 1998.
\bibitem{toronto} R. P. Beukens, Annual Report, IsoTrace Laboratory, 
Canadian Centre for
Accelerator Mass Spectroscopy at the University of Toronto. 
ISBN 0-7727-6953-2 (1992)
\bibitem{schoenert} S. Schoenert, private communication and E. Resconi, 
PhD Thesis,
Univ. of Genova, February 2001 (Chapter 3).
\bibitem{ore} A. J. T. Hull, D. L. Barker and D. J. Donahue, Chemical 
Geology 66: 35 (1987).
\bibitem{nolte} E. Nolte, private communication.
\bibitem{hunt} J. M. Hunt, Petroleum Chemistry and Geology (2nd edition), 
W. H. Freeman and Co., 1996.
\bibitem{turekian} Turekian and Wedepohl, Bull. Geol. Soc. America,
ol. 72, 175, 1961.
\bibitem{krauskopf} Introduction to Geochemistry, McGraw Hill, 1967.
\bibitem{harder} H. Harder, Handbook of geochemistry, Springer Verlag, 1974.
\bibitem{florko} T. Florkowski, Journal of Physics G: Nuclear Particle
Physics 17: S513 (1991).
\bibitem{ishida} T. Ishida, private communication.
\bibitem{revie} R. Wolfgang, Prog. React. kinet. 1965, 3, 97.
\bibitem{nine}  G. Stocklin {\it et al.}, J. Phys. Chem. 67: 1735, 1963.
\bibitem{ten} C. McKay and R. Wolfgang, J. Am. Chem.Soc. 83: 2394, 1961. 
\bibitem{eleven} G. Stocklin and A. Wolf, J. Am. Chem.Soc. 85: 229, 1963. 
\bibitem{twelve} C. E. Lang and A. F. Voigt, J. Phys. Chem. 65: 1542, 1961.
\bibitem{t13} E. P. Rack, C. E. Lang and A. F. Voigt, J.  Chem.Phys. 38: 1211, 1963.
\bibitem{oil1} Arthleth, K.H., 1968, Maine Prairie gas field, Solano County, California.  In:
Natural Gases of North America (B.W. Beebe and B.F. Curtis, eds.):  AAPG Mem.
9, p. 79-84.
\bibitem{oil2}
Burroughs, E., Beecroft, G.W.
and Barger, R.M., 1968, Rio Vista gas field, Solano, Sacramento, and Contra
Costa Counties, California.  In: Natural Gases of North America (B.W. Beebe and
B.F. Curtis, eds.):  AAPG Mem. 9, p.  93-101.
\bibitem{oil3}
Ditzler, C.C., and Vaughan, R.H.,
1968, Brentwood oil and gas field, Controa Costa County, California.  In:
Natural Gases of North America (B.W. Beebe and B.F. Curtis, eds.):  AAPG Mem.
1968, Brentwood oil and gas field, Controa Costa County, California.  In:
Natural Gases of North America (B.W. Beebe and B.F. Curtis, eds.):  AAPG Mem.
9, p. 104-112.
\bibitem{oil4}
Rudkin, G.H., 1968, Natural gas in San Joaquin Valley, California.  In Natural
Gases of North America (B.W. Beebe and B.F. Curtis, eds.):  AAPG Mem. 9, p.
113-134.
\bibitem{oil5}
Dryden, J.E., Erickson, R.C., Off, T., and Yost, S.W., 1968, Gas in Cenozoic
rocks in Ventura-Santa Maria basins, California.  In: Natural Gases of North
America (B.W. Beebe and B.F. Curtis, eds.):  AAPG Mem. 9, p. 135-148.  
\bibitem{oil6}
Osmond, J.C., Locke, R.,
Dille, A.C., Praetorious, W., and Wilkins, J.G., 1968, Natural gas in Uinta
basin, Utah.  In: Natural Gases of North America (B.W. Beebe and B.F. Curtis,
eds.):  AAPG Mem. 9, p. 174-198.
\bibitem{loomis} T. Loomis, private communication.
\end{thebibliography}
\end{document}